\documentclass[iop]{emulateapj}
\usepackage{apjfonts}
\usepackage{epsfig}
\usepackage{natbib}
\shortauthors{Frinchaboy et al.}
\shorttitle{OCCAM: Metallicity Gradient using APOGEE DR10}

\begin{document}

\title{The Open Cluster Chemical Analysis and Mapping Survey: Local
  Galactic Metallicity Gradient with APOGEE using SDSS DR10}

\author{Peter M. Frinchaboy\altaffilmark{1}, 
 Benjamin Thompson\altaffilmark{1},
 Kelly M. Jackson\altaffilmark{1,2},
 Julia O'Connell\altaffilmark{1},
 Brianne Meyer\altaffilmark{1},
 Gail Zasowski\altaffilmark{3,4,5}, 
 Steven R. Majewski\altaffilmark{6},
 S. Drew Chojnowksi\altaffilmark{6}, 
 Jennifer A. Johnson\altaffilmark{4,5},
 Carlos Allende Prieto\altaffilmark{7,8},
 Timothy C. Beers\altaffilmark{9}, 
 Dmitry Bizyaev\altaffilmark{10},
 Howard Brewington\altaffilmark{10},
 Katia Cunha\altaffilmark{11},
 Garrett Ebelke\altaffilmark{10},
 Ana Elia Garc{\'i}a P{\'e}rez\altaffilmark{6},
 Frederick R. Hearty\altaffilmark{6},
 Jon Holtzman\altaffilmark{12},
 Karen Kinemuchi\altaffilmark{10},
 Elena Malanushenko\altaffilmark{10}, 
 Viktor Malanushenko\altaffilmark{10}, 
 Moses Marchante\altaffilmark{10},
 Szabolcs M{\'e}sz{\'a}ros\altaffilmark{7,8},
 Demitri Muna \altaffilmark{4},
 David L. Nidever\altaffilmark{13},
 Daniel Oravetz\altaffilmark{10},  
 Kaike Pan\altaffilmark{10}, 
 Ricardo P. Schiavon\altaffilmark{14}, 
 Donald P. Schneider\altaffilmark{15,16}, 
 Matthew Shetrone\altaffilmark{17}, 
 Audrey Simmons\altaffilmark{10},
 Stephanie Snedden\altaffilmark{10},
 Verne V. Smith\altaffilmark{18,19},
 John C. Wilson\altaffilmark{6} 
 } 
 
%

\altaffiltext{1}{Department of Physics \& Astronomy, Texas Christian University,
TCU Box 298840, Fort Worth, TX 76129 (p.frinchaboy, b.a.thompson1,
 j.oconnell, b.r.meyer@tcu.edu)}
\altaffiltext{2}{Current Address: Department of Physics, University of Texas--Dallas,
Dallas, TX 75080, USA (Kelly.Jackson@utdallas.edu)}

\altaffiltext{3}{NSF Astronomy and Astrophysics Postdoctoral Fellow (gail.zasowski@gmail.com)}
\altaffiltext{4}{Department of Astronomy, Ohio State University, Columbus,
OH 43210 (jaj, muna@astronomy.ohio-state.edu)}
\altaffiltext{5}{Center for Cosmology and Astro-Particle Physics, Ohio
State University, Columbus, OH 43210}
\altaffiltext{6}{Department of Astronomy, University of Virginia, 
P.O. Box 400325, Charlottesville, VA 22904-4325, USA (srm4n, sdc4sb, aeg4x,
frh3z, mfs4n, jcw6z@virginia.edu)}

\altaffiltext{7}{Instituto de Astrof{\'i}sica de Canarias, 38205 La Laguna,
Tenerife, Spain (callende, meszi@iac.es)}
\altaffiltext{8}{Departamento de Astrof{\'i}sica, Universidad de La Laguna,
38206, La Laguna, Tenerife, Spain}
\altaffiltext{9}{National Optical Astronomy Observatory and JINA:
  Joint Institute for Nuclear Astrophysics, Tucson, AZ 85719, USA (beers@noao.edu)}
\altaffiltext{10}{Apache Point Observatory, P.O. Box 59, Sunspot, NM
  88349-0059, USA (dmbiz, hjbrew, gebelke, kinemuchi, elenam, viktorm,
  marchante, doravetz, kpan, asimmons, sneeden@apo.nmsu.edu)}
\altaffiltext{11}{Observat{\'o}rio Nacional, S{\~a}o
  Crist{\'o}v{\~a}o, Rio de Janeiro, Brazil (cunha@email.noao.edu)}
\altaffiltext{12}{Department of Astronomy, MSC 4500, New Mexico State
University, P.O. Box 30001, Las Cruces, NM 88003 (holtz@nmsu.edu)}
\altaffiltext{13}{Department of Astronomy, University of Michigan, Ann
  Arbor, MI, 48109, USA (dnidever@umich.edu)}
\altaffiltext{14}{Astrophysics Research Institute, Liverpool John
  Moores University, Wirral, CH41 1LD, UK (rpschiavon@gmail.com)}
\altaffiltext{15}{Department of Astronomy and Astrophysics, The Pennsylvania State University,
   University Park, PA 16802 (dps7@psu.edu)}
\altaffiltext{16}{Institute for Gravitation and the Cosmos, The Pennsylvania State University,
   University Park, PA 16802}
\altaffiltext{17}{McDonald Observatory, The University of Texas at
  Austin, Austin, TX, 78712, USA (shetrone@astro.as.utexas.edu)}
\altaffiltext{18}{National Optical Astronomy Observatories, Tucson, AZ 85719, USA (vsmith@email.noao.edu)}
\altaffiltext{19}{Steward Observatory, University of Arizona, Tucson, AZ 85721, USA}

\begin{abstract}
The Open Cluster Chemical Analysis and Mapping (OCCAM) Survey 
aims to produce a comprehensive, uniform, infrared-based dataset for
hundreds of open clusters, and constrain key Galactic dynamical
and chemical parameters from this sample.
This first contribution from the OCCAM survey 
presents analysis of 141 members stars in 28 open clusters with high-resolution metallicities derived
from a large uniform sample collected as part of
the SDSS-III/Apache Point Observatory Galactic Evolution Experiment
(APOGEE).  This sample includes the first high-resolution metallicity measurements for 22 open clusters.  
With this largest ever uniformly observed sample of open cluster stars 
we investigate the Galactic disk
gradients of both [M/H] and [$\alpha$/M].  We find basically no
gradient across this range in [$\alpha$/M], but [M/H] does show a gradient for 
$R_{GC} < 10$ kpc and a significant flattening beyond $R_{GC} = 10$ kpc.  
In particular, whereas fitting a single 
linear trend yields an [M/H] gradient of $-0.09 \pm 0.03$ dex kpc$^{-1}$ 
--- similar to previously measure gradients inside 13 kpc ---  
by independently fitting inside and outside 10 kpc separately
we find a significantly
steeper gradient near the Sun ($7.9 \le R_{GC} \le 10$) than
previously found ($-0.20 \pm 0.08$ dex kpc$^{-1}$) and a nearly flat trend
beyond 10 kpc ($-0.02 \pm 0.09$ dex kpc$^{-1}$).

\end{abstract}

\section{ INTRODUCTION }

%
A key observable used to constrain galaxy evolution models and often explored in external galaxies is the 
variation of chemical abundances across galaxy disks.  The Milky Way
provides the one galaxy where
we can study these variations in utmost detail using high resolution spectroscopy.
Open clusters have long been used as a key Galactic tracer to
probe chemical and age distributions within the Milky Way
disk (e.g., Carraro \& Chiosi 1994; Janes \& Phelps 1994)
because they provide the most
reliable `age-datable' population tracer
at low latitudes.  However, as traced by open clusters, the Galactic abundance ``gradient'' has
been fit by a single linear gradient, a 2-function gradient, a
polynomial, or a step function.
Thus, despite extensive work in this area 
(e.g., Bragaglia et al.\ 2008,
Sestito et al.\ 2008, Jacobson et al.\ 2009, Pancino et al.\ 2010; Friel
et al.\ 2010, Yong et al.\ 2012), a clear picture remains elusive, complicated by
observational limitations --- e.g., inhomogeneous datasets and small statistical samples, both 
numbers of clusters and stars per cluster (typically only 1-2 stars each).  
Yong et al.\ (2012) 
summarizes the state of the field:
{\em ``that
  definitive conclusions await homogeneous analyses of larger samples
  of stars in larger numbers of clusters. Arguably, our understanding
  of the evolution of the outer disk from open clusters is currently
  limited by systematic abundance differences between various
  studies''} 

We aim to resolve this problem by analyzing open clusters while
taking advantage of a unique, new survey, the Apache Point Observatory 
Galactic Evolution Experiment (APOGEE; Allende Prieto et al.\ 2008; 
Majewski et al.\ 2010),
one of four projects included in the Sloan Digital Sky Survey III
(SDSS-III; Eisenstein et al.\ 2011). 
APOGEE is an infrared ($H$-band) high-resolution ($R \sim 22,500$) survey of the Galaxy
that, due to its $\sim 7$ sq. deg. field of view (Gunn et al.\ 2006), will eventually target
stars in the fields of hundreds of open clusters in process of
surveying the Galaxy.  
We will leverage the strength of the APOGEE cluster
catalog to probe chemical trends in the Galactic disk
and, in particular, the behavior of the abundances as a
function of $R_{GC}$ in the transition region between the solar
neighborhood
and the outer disk, where there remains debate
about the slope and even form (e.g., linear or bilinear gradient, polynomial, or ``step function'') of 
the abundance trend 
(e.g., Corder \& Twarog
2001; Chen et al.\ 2003; Yong et al.\ 2005;  Magrini et al.\
2009; Jacobson 2009; Friel et al.
2010).


In this first contribution from the Open Cluster Chemical Analysis and Mapping survey, or ``OCCAM'' survey 
we
explore the local ($7.9 \le R_{GC} \le 14.5$) Galactic
gradients of both [M/H] and [$\alpha$/M] using data from the first of three
years of the APOGEE survey. 

\section{The Open Cluster Chemical Analysis and Mapping (OCCAM) Survey}

The OCCAM survey goals are to create a high confidence catalog of cluster
age, distance, reddening, abundances based on {\it uniform} data, and 
utilize this sample to make marked improvements to the detailed
chemical measurement of 
the Milky Way disk 
that will inform models of galaxy evolution.
The OCCAM survey
will utilize large, uniform, well-calibrated surveys 
as its
basis, starting with infrared (IR) photometry from the Two Micron
All-Sky Survey 
(2MASS; Skrutskie et al.\ 2006), the {\it Spitzer}/IRAC-based 
Galactic Legacy Infrared Mid-Plane Survey Extraordinaire programs
(GLIMPSE-1, -2, -3D, 360; Benjamin et al.\ 2003), and
the Wide-field Infrared Survey Explorer ({\em WISE}; Wright et al.,
2010), combined with spectroscopy from the SDSS-III/APOGEE survey. 
APOGEE will provide high precision radial velocities (RVs), stellar
parameters (T$_{eff}$, $\log g$, [M/H], [C/M], [N/M], [$\alpha$/M]),
and eventually detailed abundances for individual elements 
(Fe, C, N, O, Al, Si, Ca, Ni, Na, S, Ti, Mn, K).  These data sets will  
also be combined with new precision astrometric surveys as they come
available (e.g., Pan-STARRS and Gaia; Kaiser et al.\ 2010; Casertano et
al.\ 1996).  

\subsection{The APOGEE Survey and SDSS Data Release 10 }

The tenth data release of the Sloan Digital Sky Survey (Ahn et al.\
2013, DR10) provides the
first public release of APOGEE data from first light observations in May 2012
through those taken in normal survey mode until July 2012.  The spectra extracted using the APOGEE data reduction
pipeline, which also measures RVs (in DR10, with typical uncertainties of 150 m
s$^{-1}$; Nidever et al., {\it in prep}).  
The stellar parameters and abundances have been determined using the 
APOGEE Stellar Parameters and Chemical Abundances Pipeline
(ASPCAP; Garcia Perez et al., {\it in prep}).  ASPCAP is a set
of IDL routines and a FORTRAN code called
FERRE, which finds the best fit to the observed spectrum based on a
$\chi^2$ minimization from a library
of synthetic spectra
computed for a large range of stellar parameters and abundances.
The DR10 APOGEE database contains the stellar parameters 
T$_{eff}$ and $\log g$ as well as [M/H], [C/M], [N/M], and
[$\alpha$/M] from the ASPCAP matching and interpolation. 
The verification of ASPCAP was conducted by comparing its
results to those of optical high-resolution studies for stars in a set of
``calibration'' 
open and globular clusters 
(M{\'e}sz{\'a}ros et al.\ 2013).
 The ASPCAP [M/H] provided by FERRE is well-correlated with
[Fe/H], as shown in M{\'e}sz{\'a}ros et al. (2013).

\section{The OCCAM DR10 Sample}
\subsection{ Calibration Open Clusters} 

Our study includes 6 of the 10 targeted calibration open
clusters (M67, 
NGC 2158, NGC 2420, NGC 6791, NGC 6819, NGC 7789) from M{\'e}sz{\'a}ros et al.\ (2013). 
We reanalyze their stellar membership and use for them DR10 parameters consistent with those used
for the other clusters in our study (\S3.2).
The calibration stars targeted by APOGEE in these clusters are flagged
in the DR10 database by {\it
  apogee\_target2} = 10 (Zasowski et al.\ 2013).

\subsection{``Field'' Open Clusters} 

Because of the large SDSS field of view and the plan for APOGEE to observe
all Galactic populations, but with a particular focus on the Galactic disk and bulge,
the survey is targeting a large number of open clusters with a relatively
small subset of the 300 fibers available per plate.   
To utilize
these fibers efficiently, a new technique was developed to improve the chances of targeting
cluster stars in these crowded, highly contaminated, low Galactic
latitude fields.

This technique to isolate cluster stars from the general field uses
spatial information (color-magnitude comparisons within and outside of
the cataloged cluster visual radius, $R_{cl}$) combined with filtering by reddening
as derived from the Rayleigh-Jeans Color Excess (RJCE) technique
(Majewski et al.\ 2011), which derives star-by-star extinctions ($A_{K_S}$)
that can be used to remove background and foreground stars
This method takes advantage of the fact that all stars have
IR 2MASS
and {\em Spitzer}/IRAC and/or {\em WISE} photometry.  This IR photometry allows a direct 
assessment of the line-of-sight reddening to any particular star across wavelengths
where the reddening law is nearly universal.
At these wavelengths the color effects of reddening and stellar atmospheres are almost completely
separable: the long wavelength spectral energy distributions 
of stars have the same Rayleigh-Jeans shape, equivalent to saying that the
Vega-based, {\it intrinsic} colors of all stars are nearly constant
for the correct combination of filters. Thus, the
{\it observed} mid-IR colors contain information on the reddening to a
star {\it explicitly}, whereas the near IR colors contain information on the
stellar types.  The technique identifies the cluster $A_K$ by maximizing the number of stars within the cluster radius relative to the
number ``outside''  ($1 < R_{cl} \le 2$)  of the cluster radius for a
given range of $A_K$ ($\Delta A_K = 0.1$).  Given the difference of area and the sometimes non-uniform background, we
measure the ``outside'' sample in four areas (Shown in Figure 1a)
and then have to normalize the count by the area of sky covered (e.g., sq.
degrees) 

\begin{figure}[!]
\begin{center}
\epsscale{1.2}
\plotone{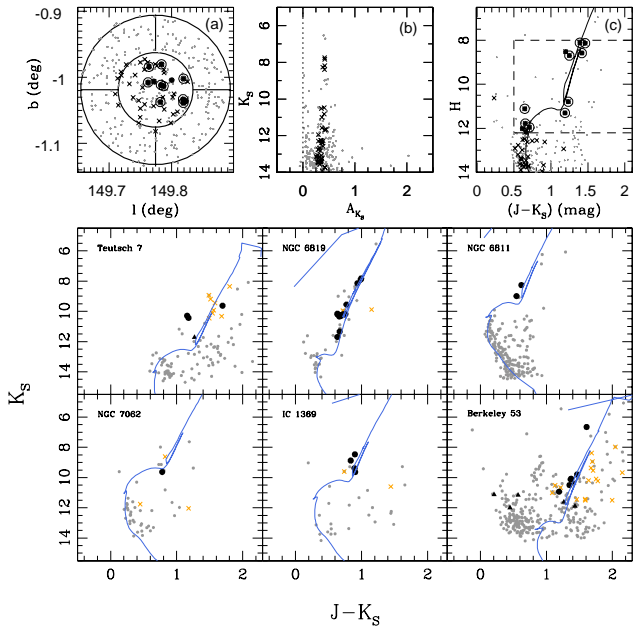}
\end{center}
\caption{ \label{fig:cl_example}    
Example analysis for the cluster King 7 using
  2MASS+WISE data. (a) Galactic longitude and latitude
  for all stars 
  within 2 cluster radii ($R_{cl}$; {\it gray points});
   stars selected to be likely members from the 
  extinction analysis are shown as {\it black crosses}.  
  Stars targeted by APOGEE are
  circled.
  (b) Distribution of $A_{K_s}$ for all stars in the King 7
  sample area; {\it black crosses} denote stars within $1.1 R_{cl}$ and within the
  associated mean cluster $A_{K_s}$ range.
  (c) Color-magnitude diagram
  (CMD) for all stars within $2R_{cl}$. The {\it dashed box}
  denotes the approximate APOGEE target selection region (however
  APOGEE uses the dereddened
  $(J-K)_0 \ge 0.5$ selection), stars within the APOGEE selection are shown as
  black squares. 
  %
  The CMD is
  overplotted with a solar metallicity Padova Isochrone
  (Marigo et al.\ 2008) using the
  cluster's parameters (age, distance, reddeneing) from Dias et al (2002).
{\it bottom}) The 2MASS color-magnitude diagrams for all open clusters in
  this study.  {\it Gray points} denote 2MASS stars within the cluster
  radius, {\it black circles} are APOGEE DR10 stars selected as members and
  having reliable [M/H] and [$\alpha$/M] measurements, and {\it black triangles}
  are RV member stars without reliable metallicity measurements. 
  {\it Orange crosses} are APOGEE stars that are non-members. 
  {\it Blue lines} are Padova isochrones (Marigo et al.\ 2008) using the
  measured APOGEE [M/H] with age, distances, and reddening from Table 1.
}
\label{fig:cl_example}    
\end{figure}
 
Stars that are isolated as above
and that lie within the APOGEE standard color-magnitude cuts ($(J-K_s)_0 \ge 0.5$ and $7 \le H \le
12.2$ for a standard 3-hr APOGEE field) can then be selected as likely
cluster members. 
For longer length APOGEE fields, fainter
stars can also be targeted (down to $H = 13.8$ for a 24-hr field). 
This
cleaning is required in most clusters for two reasons: (1) most open clusters are
at low Galactic latitude and hence are heavily contaminated
with field stars. (2) Due to the plate scale of the SDSS telescope and
the size of the fiber ferrules, the minimum fiber-to-fiber distance is fairly large ($\ge 1$ arcmin),
which allows for the targeting of only a handful of stars ($\sim
5-10$) per cluster for the most distant, reddened clusters, which tend also
to be the most poorly studied.
This method is more fully described in Frinchaboy et al., {\it
  in prep} and Zasowski et al.\ (2013).
Stars targeted by APOGEE using this method are flagged in the
DR10 database by {\it apogee\_target1} =  9 (Zasowski et al.\ 2013). 

We present a demonstration of this technique utilizing the cluster King
7, shown in Figure~\ref{fig:cl_example}.
Figure~\ref{fig:cl_example}a shows the Galactic latitude and longitude area explored by our analysis.
As described above, we selected likely cluster members utilizing 
$A_{K_s}$, as shown in Figure~\ref{fig:cl_example}b.  For King 7, we find
 a
moderate extinction to the cluster.
The color-magnitude diagram (CMD) of the cluster shown in Figure~\ref{fig:cl_example}c
highlights the member stars having $A_{K_s}$ values
within the selected window of extinction, 
and the dashed box
denotes the area of APOGEE's primary target selection ($7.0 < H < 12.2$ and $J-K_S \ge 0.5$). 
Finally, we compare our ``cleaned'' cluster CMD to the Padova 
isochrone matching catalog cluster values (Dias et al.\ 2002) for King 7
and find good agreement.  By comparing the CMD with isochrone values,
when available, we are able to isolate candidate open cluster stars with a
high probability for membership.  


\subsection{Cluster Membership and Metallicities}

To determine cluster membership, we have combined the APOGEE sample
with the UCAC-4 (Zacharias et al. 2013) dataset and have used the 3D kinematical membership
analysis from Frinchaboy \& Majewski (2008) to derive
membership probabilities for each star.
For this analysis we only use the radial velocity criterion, considering cluster members as
those stars
with radial velocity membership
probabilities $>50$\%, as many cluster lack stars with UCAC-4
proper motions.
For clusters with a single star in the cluster radius, if it was along
the Dias et al. (2002, version 3.3) catalog based isochrone fit we
assumed it was a member for this study.
After determining kinematical membership, a
3$\sigma$ iterative cut was made on the stellar metallicities to
further remove any potential non-members. 

For this paper, 
we analyze only reliable open clusters, which comprise 
clusters having member stars that clearly lie near
the cluster's CMD locus (see Figures 1-3), have survey quality data in
DR10, and have
no ASPCAP warning flags for the analyzed stars (e.g., $3500 < T_{eff} < 5500$ K and for $\log g < 3.8$). 
The clusters presented here have
a median of three cluster radial velocity members and one member star with reliable [M/H]
and [$\alpha$/M]. A fuller description and detailed analysis for all DR10 APOGEE open clusters 
will be presented in Frinchaboy et al. {\it in prep}.
Using the RV kinematic membership criteria
we analyzed 546 stars, finding 141 member stars, in 28 open clusters
whose bulk cluster parameters and
metallicities presented in Table 1. 
For 22 of these clusters, Table 1 presents
the first measured high-resolution metallicity.
This study increases the number of open clusters with published high-resolution
metallicities by about 30\%. 

After our membership census and after adopting the [M/H] derived from our analysis of the
APOGEE data, we refit the CMD by eye with Padova isochrones (Marigo
et al.\ 2008) to measure the other
fundamental cluster parameters (age, distance, and reddeneing; Figure 2, blue lines).
%
 %
For most clusters, the Dias et
al. (2002; version 3.3) catalog parameters were well fit, with only occasional
minor corrections to the reddening required.  For those clusters with a
poor fit, we updated the parameters based on our refitting. 
Stars that clearly fall off the isochrone fits (e.g., ASCC
14, NGC 1912, NGC 2234, FSR 660) were excluded from determination
of the cluster metallicity. 
The final cluster parameters used in the latter analysis
are presented in Table 1, with updated parameters shown in {\it italics}.

\begin{deluxetable*}{lrcccrcrrrrr}
\tabletypesize{\small}
\tablewidth{0pt}
\tablecaption{Selected APOGEE DR10 Open Clusters \label{cl_metals}}
\tablehead{ 
Cluster  & \multicolumn{1}{c}{Diam} &
  \multicolumn{1}{c}{$\log(Age)$\tablenotemark{a}} & \multicolumn{1}{c}{Dist\tablenotemark{a}} 
  &  \multicolumn{1}{c}{$E(B-V)$\tablenotemark{a}} &\multicolumn{1}{c}{$R_{GC}$} &\multicolumn{1}{c}{$Z_{GC}$} 
  &  \multicolumn{1}{c}{Num.} &\multicolumn{1}{c}{[M/H]}&\multicolumn{1}{c}{[$\alpha$/M]}&\multicolumn{1}{c}{New?} \\
& \multicolumn{1}{c}{(')} & \multicolumn{1}{c}{(yr)} &
  \multicolumn{1}{c}{(kpc)} &&
  \multicolumn{1}{c}{(kpc)} &
  \multicolumn{1}{c}{(kpc)}
  &  \multicolumn{1}{c}{Memb.}& \multicolumn{1}{c}{(dex)}& \multicolumn{1}{c}{(dex)}&
}
\startdata
     Teutsch 7 &  8.0  &  8.48 &      7.070  & {\it 1.33}   &  7.97  &  $-$0.07  &   3  & $ 0.17\pm0.03$ & $-0.00\pm0.06$ &  Y  \\   
      NGC 6819 &  5.0  &  9.41 &      2.432  &      0.14    &  8.18  &  $+$0.36  &  13  & $ 0.07\pm0.01$ & $ 0.01\pm0.03$ &  N \\   
      NGC 6811 & 14.0  &  8.80 &      1.215  &      0.16    &  8.36  &  $+$0.25  &   2  & $-0.02\pm0.04$ & $-0.01\pm0.07$ &  Y  \\   
      NGC 6791 & 10.0  &  9.92 &      5.035  &      0.14    &  8.29  &  $+$0.95  &  29  & $ 0.38\pm0.01$ & $ 0.13\pm0.02$ &  N  \\   
      NGC 7062 &  5.0  &  8.46 & {\it 2.100} & {\it 0.51}   &  8.75  &  $-$0.10  &   1  & $ 0.08\pm0.05$ & $-0.07\pm0.10$ &  Y  \\   
       IC 1369 &  5.0  &  8.64 &      2.083  &      0.57    &  8.74  &  $-$0.02  &   4  & $ 0.09\pm0.03$ & $-0.05\pm0.05$ &  Y  \\   
   Berkeley 53 & 22.0  &  9.09 &      3.100  & {\it 1.40}   &  9.07  &  $+$0.20  &   6  & $-0.09\pm0.06$ & $ 0.08\pm0.10$ &  Y  \\   
      NGC 7093 &  9.0  &  8.95 & {\it 3.200} & {\it 0.50}   &  9.14  &  $-$0.24  &   1  & $-0.15\pm0.13$ & $ 0.01\pm0.18$ &  Y  \\   
      NGC 2682 & 25.0  &  9.45 &      0.792  &      0.04    &  9.06  &  $+$0.42  &  21  & $ 0.01\pm0.01$ & $ 0.01\pm0.02$ &  N  \\   
      NGC 7789 & 25.0  &  9.15 &      1.795  &      0.28    &  9.41  &  $-$0.17  &  16  & $ 0.02\pm0.01$ & $-0.01\pm0.03$ &  N  \\   
 Collinder 106 & 35.0  &  9.90 &      1.000  & {\it 0.40}   &  9.41  &  $-$0.01  &   3  & $-0.26\pm0.04$ & $ 0.01\pm0.05$ &  Y  \\   
   Berkeley 91 &  3.0  &  9.35 &      4.100  & {\it 0.12}   &  9.44  &  $+$0.02  &   2  & $-0.13\pm0.04$ & $ 0.10\pm0.07$ &  Y  \\   
      FSR  498 &  1.5  &  8.55 &      1.800  & {\it 0.22}   &  9.54  &  $-$0.01  &   1  & $-0.34\pm0.07$ & $ 0.03\pm0.10$ &  Y  \\   
       ASCC 14 & 26.4  & {\it 9.00} &  1.100 & {\it 0.60}   &  9.59  &  $-$0.02  &   1  & $-0.10\pm0.03$ & $ 0.02\pm0.06$ &  Y  \\   
      NGC 1912 & 20.0  &  8.50 & {\it 1.100} & {\it 0.35}   &  9.59  &  $+$0.01  &   1  & $-0.38\pm0.04$ & $-0.01\pm0.06$ &  Y  \\   
      NGC 2240 & 11.0  &  9.20 &      1.551  &      0.04    & 10.02  &  $+$0.32  &   1  & $ 0.07\pm0.05$ & $ 0.00\pm0.10$ &  Y  \\   
        King 5 & 14.2  &  9.10 &      2.230  &      0.67    & 10.38  &  $-$0.17  &   4  & $-0.16\pm0.03$ & $ 0.00\pm0.05$ &  Y  \\   
      FSR  942 &  8.0  &  9.00 &      2.000  & {\it 0.80}   & 10.44  &  $-$0.13  &   1  & $-0.29\pm0.06$ & $-0.04\pm0.10$ &  Y  \\   
        King 7 &  7.0  &  8.82 &      2.200  &      1.25    & 10.46  &  $-$0.04  &   1  & $-0.17\pm0.06$ & $-0.01\pm0.10$ &  Y  \\   
      NGC 2420 &  5.0  &  9.30 &      2.480  &      0.06    & 10.78  &  $+$0.83  &  11  & $-0.23\pm0.02$ & $ 0.00\pm0.03$ &  N  \\   
       ASCC 15 & 24.0  &  8.60 & {\it 2.400} & {\it 0.40}   & 10.88  &  $+$0.00  &   1  & $ 0.26\pm0.05$ & $ 0.02\pm0.10$ &  Y  \\   
      FSR  821 &  9.9  &  8.80 & {\it 2.400} & {\it 0.85}   & 10.90  &  $-$0.01  &   1  & $-0.21\pm0.07$ & $-0.06\pm0.10$ &  Y  \\   
       NGC 136 &  4.0  &  8.40 &      5.220  &      0.70    & 12.03  &  $-$0.12  &   1  & $-0.05\pm0.06$ & $ 0.02\pm0.10$ &  Y  \\   
      NGC 2234 &  8.0  &  7.70 &      4.800  & {\it 1.00}   & 13.18  &  $+$0.24  &   1  & $-0.33\pm0.06$ & $ 0.01\pm0.10$ &  Y  \\   
      FSR  660 &  2.6  &  9.22 &      5.126  & {\it 0.79}   & 13.19  &  $-$0.05  &   1  & $-0.49\pm0.04$ & $ 0.01\pm0.06$ &  Y  \\   
      FSR  542 &  8.0  & {\it 8.50} & {\it 6.600}  & {\it 1.10}   & 13.53  &  $+$0.04  & 1  & $ 0.07\pm0.06$ & $ 0.07\pm0.10$ &  Y  \\   
      NGC 2158 &  5.0  &  9.02 &      5.071  & {\it 0.50}   & 13.55  &  $+$0.16  &  11  & $-0.16\pm0.02$ & $-0.01\pm0.03$ &  N  \\   
      FSR  941 & 12.0  &  8.70 &      5.800  & {\it 0.70}   & 14.17  &  $-$0.08  &   1  & $-0.21\pm0.07$ & $-0.05\pm0.10$ &  Y  \\[-1ex]   
\enddata
\tablenotetext{a}{Dias et al.\ (2002) catalog (version 3.3 -
  jan/10/2013). {\it Italics} denote refitted values and that differ
  from Dias et al.\\\\}
\end{deluxetable*}

\begin{figure}[!]
\begin{center}
\epsscale{1.1}
\plotone{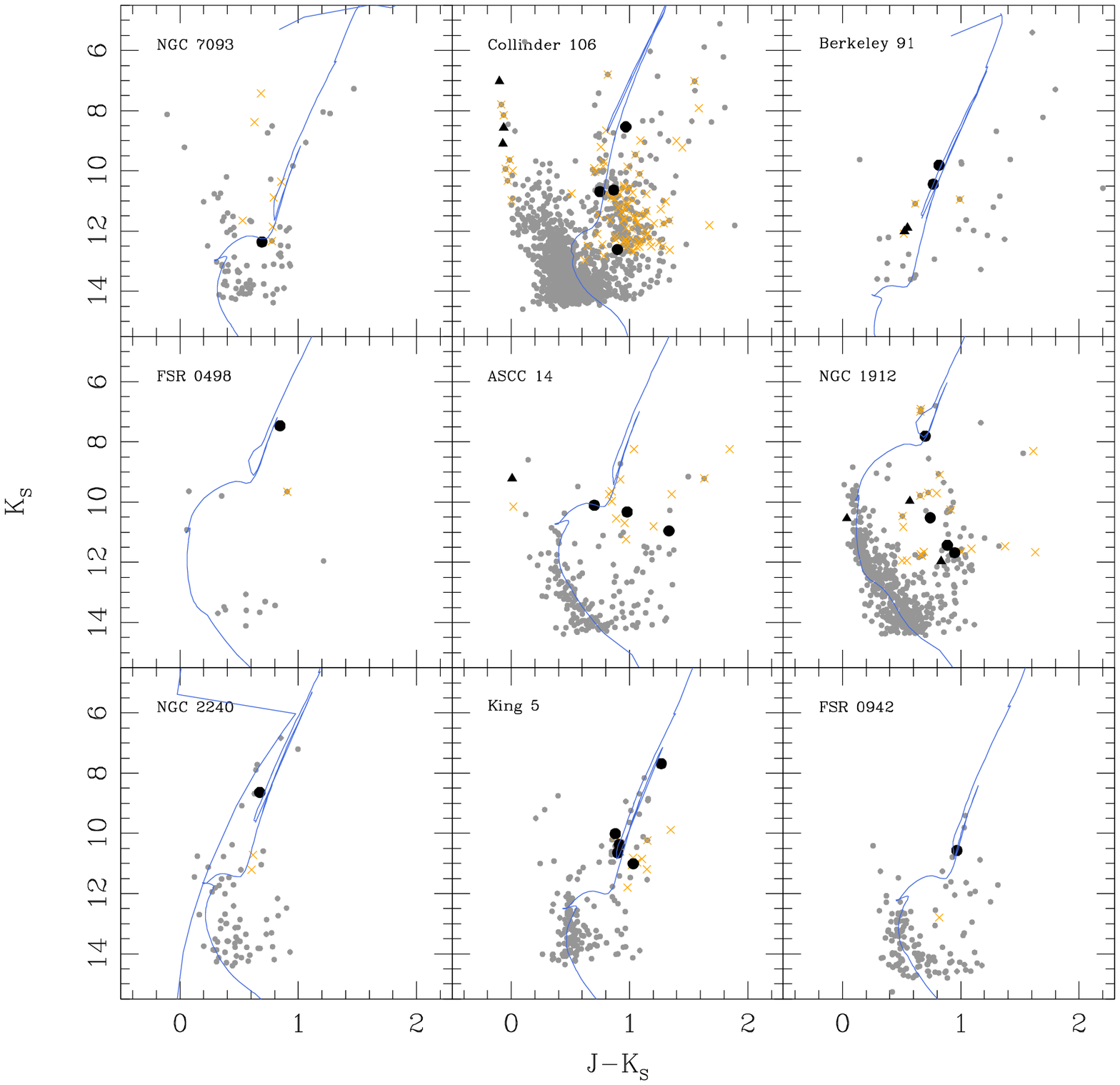}
\end{center}
\caption{Same as Figure 1 {\it bottom}} 
\end{figure}

\begin{figure}[!]
\begin{center}
\epsscale{1.1}
\plotone{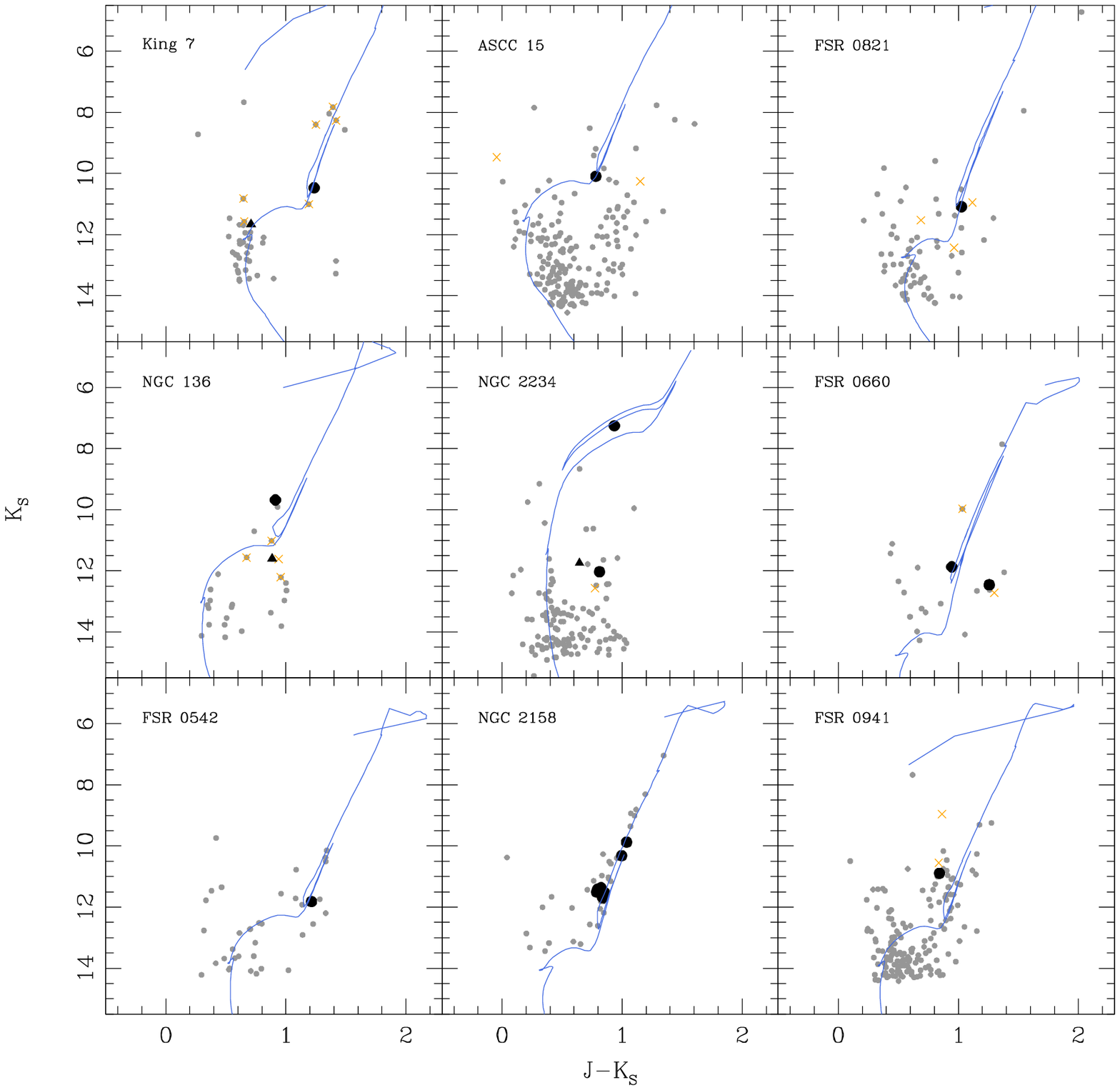}
\end{center}
\caption{Same as Figure 1 {\it bottom}} 
\end{figure}

\section{The Galactic Disk Abundance Gradient}

Using the large, uniform APOGEE sample, we explore 
Galactic abundance gradients
near and outside the solar circle ($7.9 \le R_{GC} \le 14.5$; we assume
$R_{Sun} = 8.5$ kpc).  All clusters in the sample
have $|Z| < 500$ pc, except NGC 2420 ($Z = 830$ pc) and NGC
6791 ($Z = 950$ pc).  
In this contribution, we analyze the Galactic trends
in metallicity ([M/H]) and $\alpha$-abundances ([$\alpha$/M]) only.

\subsection{[M/H] Trends}

Using distances and metallicities from Table 1, 
we assess Galactic disk metallicity trends, as shown in Figure 4.
Fitting a linear trend to our full sample, we find a gradient of 
$-0.09 \pm 0.03$ dex kpc$^{-1}$ (see Figure 4 {\it top}), 
which is steeper than, but consistent 
within the errors of previous
gradients derived from literature compilations of open clusters, e.g., $-0.06 \pm 0.02$ dex
kpc$^{-1}$ found by Pancino et al.\ (2010), Friel et al.\ (2002), and
Friel et al.\ (2012), and is closer to the gradient found by Yong et al
(2012; $-0.09 \pm 0.01$ dex kpc$^{-1}$ ).

We also investigated an alternate fit by
 independently fitting inside and outside $R_{GC}$ = 10 kpc, which is near the
 dynamical signature for Galactic co-rotation (L{\'e}pine et al.\ 2013)).
This yields a significantly
steeper gradient ($-0.20 \pm 0.08$ dex kpc$^{-1}$) near the Sun ($7.9 \le R_{GC} \le 10$ kpc) than
previously reported
(Figure 3a). 
This inner gradient is steeper than
those previously found, even for ``broken''/multi-linear fits --- e.g., 
Yong et al.\ (2012) measured a gradient of $-0.09 \pm 0.01$ dex
kpc$^{-1}$ for $R_{GC} < 13$ kpc, a similar trend to that from our full sample fit,
which covers a similar $R_{GC}$ range.

We find little gradient in our sample beyond $R_{GC} > 10$ kpc
($-0.02 \pm 0.09$ dex kpc$^{-1}$), where a plateau of
$\langle$[M/H]$\rangle = -0.3$ appears.   This plateau is similar to the flattening found by
Yong et al.\ (2012) for
clusters beyond $R_{GC} = 12$ kpc.

\begin{figure}[!]
\begin{center}
\epsscale{1.1}
\plotone{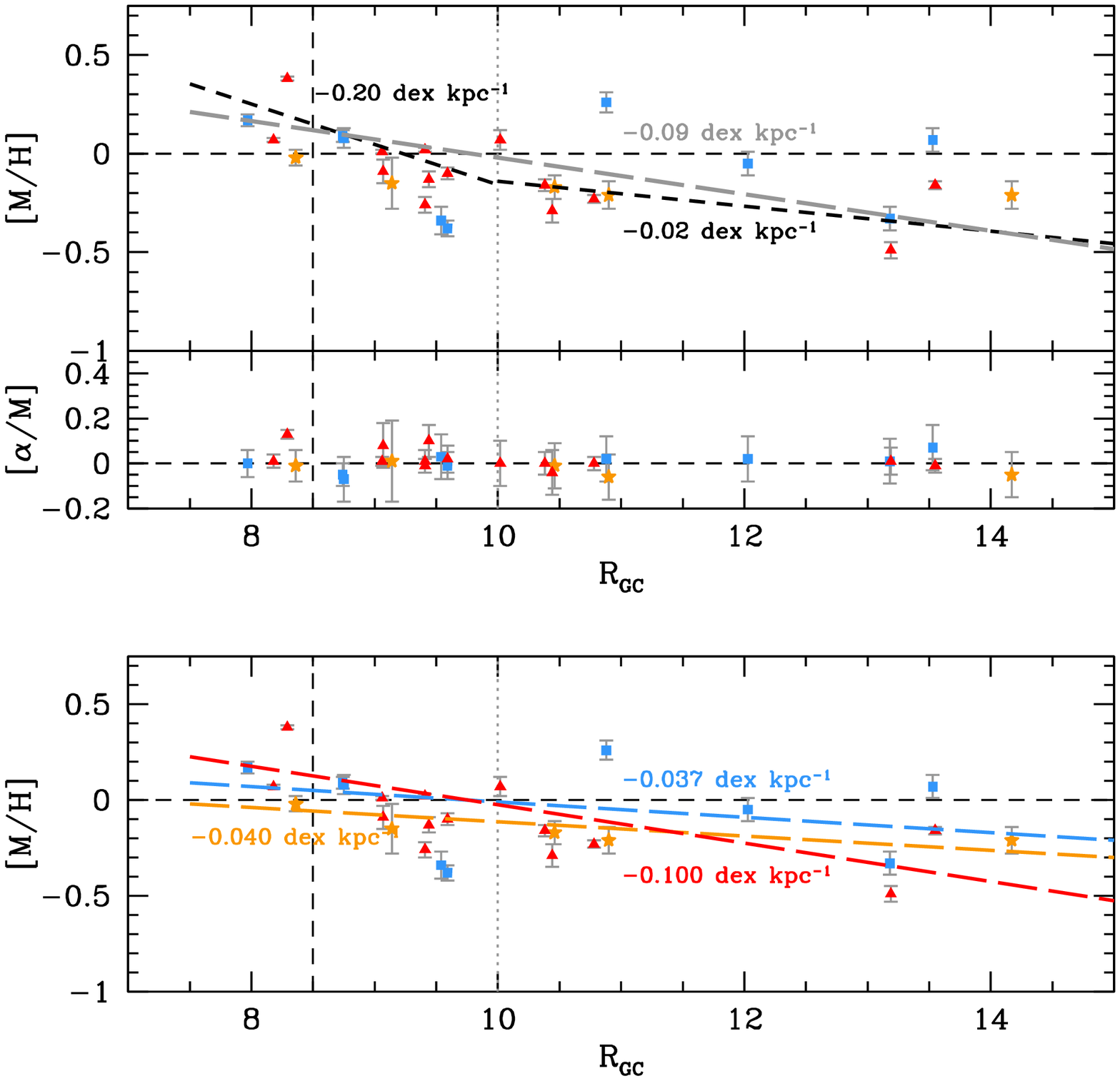}
\end{center}
\caption{ 
({\it top}) The [M/H] disk abundance gradient using APOGEE DR10
  measurements of open
  clusters, assuming $R_{\odot} = 8.5$ kpc.  A linear fit to the full sample ({\it grey dashed line}) yields a
  gradient of $-0.09 \pm 0.03$ dex kpc$^{-1}$, which is consistent with the
  results of Yong et al.\ (2012).  An alternate representation ({\it black dashed line}) is a bilinear fit broken
  at $R_{GC} = 10$ kpc: 
  we find a gradient of $-0.20 \pm 0.08$ dex kpc$^{-1}$ for the inner cluster sample ($7.9 \le R_{GC}
  < 10$ kpc)  and an
  almost flat 
  trend ($-0.02 \pm 0.09$ dex kpc$^{-1}$) for the outer clusters ($R_{GC} > 10$ kpc).
  Clusters are color-coded by age: 
  {\it blue squares} for $\log(age) < 8.5$,
  {\it orange stars} for $8.5 \le \log(age) < 9.0$, and 
  {\it red triangles} for $\log(age) \ge 9.0$. 
 ({\it middle}) The [$\alpha$/M] disk abundance gradient using APOGEE DR10 open
  clusters.  There is no significant trend in [$\alpha$/M]  ($-0.01 \pm 0.05$ dex kpc$^{-1}$) for open
  clusters spanning our current sample range of $7.9 \le R_{GC} < 14$ kpc.
 ({\it bottom}) Age gradient fits.
}
\end{figure}

\subsection{[$\alpha$/M] Trends}

We find little to no trend in [$\alpha$/M] ($-0.01 \pm 0.05$ dex kpc$^{-1}$) from analysis
of the APOGEE DR10 open cluster sample across the full range of
Galactic radius ($7.9 \le R_{GC} \le 14.5$ kpc, Figure 4 {\it middle}).  While the sample does not yet
include the most distant clusters (e.g., Berkeley 29 and Saurer 1)
that have been shown to be $\alpha$-enhanced, we should be able to
place further constraints on this trend with data from future APOGEE
data releases, which target more distant clusters.

\subsection{Time Evolution of the Gradient?}

The large age span of our sample allows us to explore dependencies of the gradients with age.
We divided the sample into three
age bins: nine ``young'' clusters ($\log(age) < 8.5$), five ``intermediate''
clusters ($8.5 \le \log(age) < 9.0$), and fourteen ``old'' clusters
($\log(age) \ge 9.0$; see Figure 4 {\it bottom}).
We fit the
gradient as a function of $R_{GC}$ using only clusters from each age
bin and find similar behavior for the 
``young'' ($\Delta$[M/H] = $-0.04 \pm 0.09$ dex kpc$^{-1}$)  and 
``intermediate'' ($\Delta$[M/H] = $-0.04 \pm 0.15$ dex kpc$^{-1}$) samples. 
The ``young'' sample is small and shows a large scatter, which could
be intrinsic, or given that
APOGEE is tuned to giant stars these could be due to less reliable parameters
(e.g. the star in FSR 542 is near the $\log g$ warning limit in DR10)
or that the one star selected could actually be a non-member.
The ``old'' sample shows a
slightly steeper gradient  ($\Delta$[M/H] = $-0.10 \pm 0.04$ dex kpc$^{-1}$), which hints that there may be some
evolution as a function of age --- a shallowing of the gradient with
time.  We caution against significant
interpretation given the limited $R_{GC}$ sampling of the present study.
With a larger sample from future APOGEE data releases, we should be able to
place tighter constraints on the age trends in the Galactic disk.

\section{Conclusions }

We describe the first results from the OCCAM collaboration's exploration of the
SDSS-III/APOGEE open cluster data as presented in DR10.  
\begin{enumerate}
\item The SDSS-III/APOGEE DR10 dataset contains
the largest sample of uniformly observed and analyzed
  high-resolution metallicity measurement of stars in open clusters,
  (141 member stars in 28 clusters), including the first high-resolution
  metallicity measurements for 22 open clusters. 
\item For this sample of clusters, spanning $7.9 < R_{GC} < 10$ kpc we find a steep inner [M/H] gradient,
  $-0.20 \pm 0.08$ dex kpc$^{-1}$, 
  but a basically flat trend ($-0.02 \pm 0.09$ dex kpc$^{-1}$)
  beyond $R_{GC} = 10$ kpc.  This inner gradient is steeper than found by
  previous measurements (Friel et al.\ 2010, Yong et al.\ 2012).  The
  location of the split in these two samples is
  near the Galactic co-rotation radius, similar to the break in
  abundance gradient found
  by L{\'e}pine et al.\ (2013) using Cepheids.
  This feature may provide a useful
  constraint for studies of galaxy evolution.

\item We find no significant [$\alpha$/M] trend over $7.9 < R_{GC} <
  14.5$ kpc.  However, other studies that have found a trend toward the outer
  disk (e.g., Yong et al. 2012) have included more distant clusters
  than presented here (e.g., Berkeley 29 and Saurer 1 at $R_{GC} \sim 20$ kpc).
\end{enumerate}

Future OCCAM publications will present the full, detailed analysis of the
entire DR10 open cluster sample, including a reanalysis of the clusters'
fundamental parameters (age, distance, and reddening) and additional chemical elements anticipated by the 
full APOGEE data release DR12 (December 2014).

\acknowledgements

PMF acknowledges funding from the TCU RCAF and JFSRP programs.
KJ acknowledges funding from a TCU SERC grant. 
KC acknowledges support for this research from the
National Science Foundation (AST-0907873).
Funding for SDSS-III has been provided by the Alfred P. Sloan
Foundation, the Participating Institutions, the NSF, and the
U.S. Department of Energy Office of Science. The SDSS-III web site is
http://www.sdss3.org/.

SDSS-III is managed by the Astrophysical Research Consortium for the
Participating Institutions of the SDSS-III Collaboration including the
University of Arizona, the Brazilian Participation Group, Brookhaven
National Laboratory, University of Cambridge, Carnegie Mellon
University, University of Florida, the French Participation Group, the
German Participation Group, Harvard University, the Instituto de
Astrofisica de Canarias, the Michigan State/Notre Dame/JINA
Participation Group, Johns Hopkins University, Lawrence Berkeley
National Laboratory, Max Planck Institute for Astrophysics, New Mexico
State University, New York University, Ohio State University,
Pennsylvania State University, University of Portsmouth, Princeton
University, the Spanish Participation Group, University of Tokyo,
University of Utah, Vanderbilt University, University of Virginia,
University of Washington, and Yale University.
\facility{Sloan}


\begin{thebibliography}{}
\bibitem[Ahn et al.(2013)]{dr10} Ahn, C.P., Anderson,
  S.F., Anderton, T., et al. 2013, \apj, {\it submitted}

\bibitem[Allende-Prieto et al.(2008)]{apogee1a} Allende Prieto, C., Majewski,
  S. R., Schiavon, R., et al.  2008, AN, 329, 1018  

\bibitem[Benjamin et al.(2003)]{glimpse} Benjamin, R.~A., Churchwell,
  E., Babler, B.L., et al. 2003, \pasp, 115, 953

\bibitem[Bragaglia et al.(2008)]{bocce1} Bragaglia A., Sestito, P.,
  Villanova, S., et al. 2008, \aap, 480, 79

\bibitem[Carraro \& Chiosi(1994)]{cc94}  Carraro, G. \& Chiosi C.  1994, \aap, 287, 761 

\bibitem[Casertano et al.(1996)]{gaia} Casertano, S., Lattanzi, M. G.,
  Perryman, M. A. C., Spagna, A., \apss, 241, 89

\bibitem[Chen et al.(2003)]{chen03} Chen, L., Hou, J.L., Wang, J.J. 2003, \aj, 125, 1397

\bibitem[Corder \& Twarog(2001)]{corder01}  Corder, S. \& Twarog,
  B.A. 2001, \aj , 122, 895 

\bibitem[Dias et al.(2002)]{dias02} Dias, W.S., Alessi, B.S., Moitinho, A.,
  L\'epine, J.R.D. \& Alessi, B.S. 2002, \aap, 389, 8718 

\bibitem[Eisenstein et al.(2011)]{sdss3} Eisenstein, D.J., Weinberg, D.H., Agol, E., et al.  2011, \aj, 142, 72

\bibitem[Friel et al.(2002)]{friel02} Friel, E.D., Janes, K.A.,
  Tavarez, M., et al.\ 2002, \aj, 124, 2693

\bibitem[Friel et al.(2010)]{friel10} Friel, E.D., Jacobson, H.R., Pilachowski,
  C.A. 2010, \aj, 139, 1942

\bibitem[Frinchaboy \& Majewski(2008)]{fm08}  Frinchaboy, P.M. \& Majewski, S. R.  2008, \aj,  136, 188  



\bibitem[Gunn et al.(2006)]{gunn07} Gunn, J.E., Siegmund, W.A.,
  Mannery, E.J., et al.  2006, \aj, 131, 2332 

\bibitem[Harris(1996)]{gccat} Harris, W.E. 1996, \aj, 112, 1487 

\bibitem[Jacobson et al.(2009)]{jacob09} Jacobson, H.R., Friel, E.D., Pilachowski,
  C.A. 2009, \aj, 137, 4753

\bibitem[Janes \& Phelps(1994)]{jp94} Janes, K.A. \& Phelps.R.L.  1994, \aj, 108, 1773

\bibitem[Kaiser et al.(2010)]{panstarrs} Kaiser, N., Burgett, W.,
  Chambers, K., et al.  2010, SPIE, 7773, 12 

\bibitem[L{\'e}pine et al.(2013)]{lp13} L{\'e}pine, J.R.D., Andrievky,
  S., Barros, D.A., Junqueira, T.C., Scarano, S.  2013,  Proceedings
  of IAU Symposium 298 (arXiv:1307.7781) 

\bibitem[Magrini et al.(2009)]{bocce3} Magrini, L., Sestito, P., Randich, S., Galli,
  D. 2009, \aap, 494, 95

\bibitem[Majewski et al.(2010)]{apogee1b}
Majewski, S.R., Wilson, J.C., Hearty, F., Schiavon, R.R., Skrutskie,
M.F.  2010, IAU Symposium, 265, 480

\bibitem[Majewski et al.(2011)]{RJCE} Majewski, S.R., Zasowski, G., \& Nidever,
  D.L.  2011, \apj, 

\bibitem[M{\'e}sz{\'a}ros et al.(2012)]{apmodel} M{\'e}sz{\'a}ros, Sz., Allende
  Prieto, C., Edvardsson, B., et al.  2012, \aj, 144, 120 

\bibitem[M{\'e}sz{\'a}ros et al.(2013)]{apcalib}  M{\'e}sz{\'a}ros, Sz.,
  Holtzman, J., Garc{\'i}a P{\'e}rez, A.E., et al. 2013, \aj, {\it submitted}


\bibitem[Pancino et al.(2010)]{p10}  
Pancino, E., Carrera, R., Rossetti, E., Gallart, C.   2010, \aap, 511, A56

\bibitem[Sestito et al.(2008)]{bocce2} Sestito, P., Bragaglia, A., Randich,
  S., et al. 2006, \aap, 458, 121 

\bibitem[Skrutskie et al.(2006)]{2mass} Skrutskie, M.~F., Cutri,
  R. M., Stiening, R., et al.  2006, \aj, 131, 1163

\bibitem[Wright et al.(2010)]{wise} Wright, E.L., Eisenhardt, P.R.M., Mainzer, A., et al. 2010, \aj, 140, 1868

\bibitem[Yong et al.(2005)]{yong05} Yong, D., Carney, B. \& de Almeida, M.L.T.  2005,
  \aj, 130, 597  

\bibitem[Yong et al.(2012)]{yong12} Yong, D., Carney, B.W., Friel,
  E.D.  2012, \aj, 144 95

\bibitem[Zachariaset al.(2013)]{ucac4}  Zacharias, N., Finch, C.T.,
  Girard, T.,M., et al. 2013, \aj, 145, 44 

\bibitem[Zasowski et al.(2013)]{aptarg} Zasowski, G., Johnson, J.A.,
  Frinchaboy, P.M., et al. 2013, \aj, {\it in press}

\end{thebibliography}
\end{document}